\documentclass[secthm,seceqn,amsthm,ussrhead,eqno]{amsart}
\usepackage{amsmath,latexsym}

\usepackage[psamsfonts]{amssymb}
\usepackage{times}
\usepackage[mathcal]{euscript}
\numberwithin{equation}{section}

\newcommand{\bbT}{\mathbb T}

\renewcommand{\epsilon}{\varepsilon}

\newcommand{\be}{\begin{equation}}
\newcommand{\ee}{\end{equation}}
\newcommand{\no}{\nonumber}

\newcommand{\C}{\mathbb{C}}

\newcommand{\N}{\mathbb{N}}

\newcommand{\R}{\mathbb{R}}
\renewcommand{\S}{\mathbb{S}}
\newcommand{\T}{\mathbb{T}}

\newcommand{\Z}{\mathbb{Z}}

\newcommand{\cB}{{\mathcal B}}

\newcommand{\cI}{{\mathcal I}}

\newcommand{\cR}{{\mathcal R}}

\newcommand{\cU}{{\mathcal U}}

\newcommand{\const}{\mathrm{const}}

\renewcommand{\det}{\mathop{\mathrm{det}}}

\newcommand{\Img}{\mathop{\mathrm{Im}}}

{\bf}{\it}
\newtheorem{theorem}{Theorem}[section]
\newtheorem{lemma}[theorem]{Lemma}

\newtheorem{corollary}[theorem]{Corollary}
\newtheorem{hypothesis}[theorem]{Hypothesis}

\newtheorem{definition}[theorem]{Definition}

\newtheorem{remark}[theorem]{Remark}

\date{\today}

\begin{document}
\title[On the number of eigenvalues of a model operator...]
{On the number of eigenvalues of a model operator associated to a
system of three-particles on lattices}

\author{Sergio  Albeverio$^{1,2,3}$, Saidakhmat  N.~Lakaev$^{4,5}$,
  Zahriddin I.~Muminov $^{5}$}

\address{$^1$ Institut f\"{u}r Angewandte Mathematik,
Universit\"{a}t Bonn, Wegelerstr. 6, D-53115 Bonn\ (Germany)}

\address{
$^2$ \ SFB 611, \ Bonn, \ BiBoS, Bielefeld - Bonn;}
\address{
$^3$ \ CERFIM, Locarno and Acc.ARch,USI (Switzerland) E-mail:
albeverio@uni.bonn.de}

\address{
{$^4$ Samarkand State University, University Boulevard 15, 703004,
Samarkand (Uzbekistan)} \ {E-mail: slakaev@mail.ru }}

\address{$^5$ Samarkand division of Academy of sciences of
Uzbekistan (Uzbekistan) E-mail:~zimuminov@mail.ru }

\begin{abstract}
A model operator $H$ associated to a system of three-particles on
the three dimensional lattice $\Z^3$ and interacting via  pair
non-local potentials is studied. The following results are proven:
(i) the operator $H$ has  infinitely many eigenvalues lying below
the bottom of the essential spectrum and accumulating at this point,
in the case, where  both Friedrichs model operators
$h_{\mu_\alpha}(0),\alpha=1,2,$ have threshold resonances. (ii) the
operator $H$ has a finite number of eigenvalues lying outside of the
essential spectrum, in the case, where at least one
 of
 $h_{\mu_\alpha}(0),\,\alpha=1,2,$  has a threshold
eigenvalue.

\end{abstract}
 \maketitle



\section{Introduction}

The main goal of the present paper is to prove the finiteness or
infiniteness of the number of eigenvalues for a model operator $H$
with emphasis on the asymptotics for the number of infinitely many
eigenvalues (Efimov's effect case).
 The model
operator $H$ is associated to a system of three-particles on the
lattice $\Z^3$ interacting via  pair non-local potentials.

The Efimov  effect is one of the most remarkable results in the
spectral analysis for continuous three-particle  Schr\"{o}dinger
operators: if none of the three two-particle Schr\"{o}dinger
operators (corresponding to the two-particle subsystems) has
negative eigenvalues, but at least two of them have a zero energy
resonance, then this three-particle Schr\"{o}dinger operator has an
infinite number of discrete eigenvalues, accumulating at zero.

Since its discovery by Efimov in \cite{Efi73} many works have been
devoted to this subject. See, for example
\cite{AHW,AmNo,DFT,FaMe,OvSi,Sob93,Tam91,Tam94,Yaf74}.

The main result obtained by  Sobolev \cite{Sob93} (see also
\cite{Tam94}) is an asymptotics of the form $\cU_0|log|\lambda||$
for the number $N(\lambda)$ of eigenvalues on the left of
$\lambda,\lambda<0$, where the coefficient ${\cU}_0$ does not depend
on the two-particle potentials $ v_\alpha $ and is a positive
function of the ratios $m_1/m_2,m_2/m_3$ of the masses of the
three-particles.

In models of solid state physics
\cite{GrSc,FIC,Mat86,Mog91,ReSiIII,Yaf00} and also in lattice
quantum field theory \cite{MaMi} discrete Schr\"{o}dinger operators
are considered, which are lattice analogues of the three-particle
Schr\"{o}dinger operator in a continuous space. The presence of the
Efimov effect for these operators was proved in
\cite{ALzM04,Lak91,Lak93}.

In \cite{ALzM04} a system of three arbitrary quantum particles on
the lattice $\Z^3$ interacting via zero-range pair attractive
(local) potentials is considered and for the number of eigenvalues
$N(\lambda)$ an asymptotics analogous to \cite{Sob93,Tam94}  have
been obtained.

In all papers devoted to Efimov's effect   systems of three
particles interacting via pair local potentials have been considered.

In the present paper we study the model operator $H$ associated to a
system of three-particles on  $\Z^3$
 acting in the Hilbert space $L_2(({\T}^3)^2)$ and  interacting via
pair non-local potentials, where the role of the two-particle
discrete Schr{\"o}dinger operators is  played by a family of
Friedrichs  models with parameters
$h_{\mu_\alpha}(p),\,\alpha=1,2,\,p \in \T^3.$

Under some natural conditions on the family of the operators
$h_{\mu_\alpha}(p),\,\alpha=1,2,\,p \in \T^3,$
 we obtain the following results:

(i) The essential spectrum of it is described via the spectr  of
the Friedrichs models $h_{\mu_\alpha}(p),$ $\alpha=1,2,$ $p \in
(-\pi,\pi]^3.$

(ii) the operator $H$ has infinitely many eigenvalues lying below
the bottom
  and accumulating at the bottom of its essential spectrum, in the
  case,
  where  both operators
$h_{\mu_\alpha}(0),\alpha=1,2,$ have   threshold energy resonances.
Moreover for the number $N(z)$ of eigenvalues of $H$ lying below
$z<m=\inf\sigma_{ess}(H)$ the following limit exists
\begin{equation*}
\lim\limits_{z \to m-0}\frac{N(z)}{|\log |m-z||}={\cU}_0 \quad
(0<{\cU}_0 <\infty).
\end{equation*}
(iii) the operator $H$ has a finite number of eigenvalues lying
below the bottom of its essential spectrum in the case,where at
least one of the Friedrichs  models
$h_{\mu_\alpha}(0),\,\alpha=1,2,$ has a threshold eigenvalue.

We remark that the assertion (ii) is similar to the case of the
three-particle continuous and discrete Schr\"odinger operators and
the assertion (iii) is surprising. Similar assertions  do not seem
to have been yet proved  for the three-particle Schr\"odinger operators on
$\R^3$ and $\Z^3.$

The plan of  this paper is as follows:

 Section 1 is an
introduction to the whole work. In Section 2 the model operator $H$
is introduced and the main results of the present paper are
formulated. In Section 3 we recall concepts and results concerning
the threshold analysis of the family of  Friedrichs models. In
section 4 we study the location of the essential spectrum and prove
a realization of the Birman-Schwinger principle for $H.$ The
finiteness of the number of eigenvalues of the operator $H$ is
proved in Section 5. In Section 6 an asymptotic formula for the
number of eigenvalues of $H$ is obtained.

Throughout the present paper we adopt the following conventions: The
subscript $\alpha$ (and also $\beta$) always is equal to $1$ or $2$
and $\alpha\neq\beta$ and $\T^3$ denotes  the three-dimensional
torus, the cube $(-\pi,\pi]^3$ with appropriately  identified sides.
For each $\delta>0$ the notation $U_{\delta}(0) =\{p\in
{\bbT}^3:|p|<\delta \}$ stands for a $\delta$-neighborhood of the
origin.

Denote by $L_2(\Omega)$ the Hilbert space of square-integrable
functions defined on a measurable set $\Omega \subset \R^n,$ and by
$L_2^{(2)}(\Omega)$ the Hilbert space of two-component vector
functions $f=(f_1,f_2),$ $f_\alpha \in L_2(\Omega),\,\alpha=1,2.$ We
denote by $diag\{B_1,B_2\}$ the $2\times 2$  diagonal matrix  with
operators $B_1,B_2$ as diagonal entries.

 Let  $\cB (\theta,\T^3)$ with
$1/2<\theta\leq 1$, be the Banach spaces of H\"older continuous
functions on ${\bbT}^3$ with exponent $\theta$ obtained by the
closure of the space of smooth (periodic) functions $f$ on
${\bbT}^3$ with respect to the norm
\begin{equation*}
\|f\|_{\theta}=\sup_{t, \ell\in {\bbT}^3 \atop \ell\neq 0 }\bigg
[|f(t)|+|\ell|^{-\theta}|f(t+\ell)-f(t)|\bigg ].
\end{equation*}

The set of functions $f: {\bbT}^3\to \R$ having continuous partial
derivatives up to order $n$ inclusive will be denoted
$C^{(n)}({\bbT}^3).$ In particular $C^{(0)}({\bbT}^3)=C({\bbT}^3)$
by our convention that $f^{(0)}(x)=f(x).$

\section{Three particle model operator and
statement of the  results}

Let us consider the operator $H$ acting in the Hilbert space
$L_2((\T^3)^2)$ by
\begin{equation}\label{oper H}
H=H_0-\mu_1V_1-\mu_2V_2,
\end{equation}
where
\begin{equation*}
(H_0f)(p,q)=u(p,q)f(p,q), \quad f\in L_2((\T^3)^2)
\end{equation*}
and $V_{\alpha},\alpha=1,2,$ are non-local interaction operators
\begin{align*}&(V_1f)(p,q)=\varphi_1(p)\int_{{\T}^3}\varphi_1(t) f(t,q)dt, \quad
f\in L_2((\T^3)^2),\\
&(V_2f)(p,q)=\varphi_2(q)\int_{{\T}^3}\varphi_2(t) f(p,t)dt, \quad
f\in L_2((\T^3)^2).
\end{align*}

Here $u$ is a real-valued
 essentially bounded function on $({\T}^3)^2$ and $\varphi_\alpha,$
  $\alpha=1,2,$ are
 real-valued functions and belong to $L_2(\T^3)$   and
$\mu_{\alpha},\alpha=1,2,$ are positive  real numbers.

Under these assumptions the operator $H$ defined by \eqref{oper H}
is bounded and self-adjoint.

Throughout this paper we assume  the following additional
hypotheses.

\begin{hypothesis}\label{hypoth} $(i)$ The function $u$ is even on
$({\T }^3)^2$ with respect to $(p,q),$ and has a unique
non-degenerate
 minimum at the point $(0,0)\in ({\T}^3)^2$ and all third order
partial derivatives of  $u$ belong to $\cB (\theta,(\T^3)^2), \,
\frac{1}{2}<\theta
\leq 1.$\\
$(ii)$ For some positive definite matrix $U$ and real numbers
$l,l_1, l_2 \,(l_1,l_2>0,l\not=0)$ the following equalities hold
$$
\left( \frac{\partial^2 u(0,0)}{\partial p^{(i)} \partial p^{(j)}}
\right)_{i,j=1}^3= l_1 U,\,\, \left( \frac{\partial^2
u(0,0)}{\partial p^{(i)} \partial q^{(j)}} \right)_{i,j=1}^3= l
U,\,\, \left( \frac{\partial^2 u(0,0)}{\partial q^{(i)} \partial
q^{(j)}} \right)_{i,j=1}^3= l_2 U.$$
\end{hypothesis}

\begin{remark}
The function $u$ is even and has a unique non-degenerate minimum on
$\T^3$ and hence without loss of generality we have assumed that the
function $u$ has a unique minimum at the point $(0,0)\in (\T^3)^2.$
\end{remark}
\begin{hypothesis}\label{hyp.varphi} The function
 $\varphi_{\alpha}\in C^{(2)}(\T^3),$
 $\alpha=1,2,$  is either even or odd on $\T^3.$
\end{hypothesis}

Set
$$
u_p^{(1)}(q)=u(q,p),\quad u_p^{(2)}(q)=u(p,q).
  $$

To study  spectral properties of the operator $H$ we introduce the
following two families of bounded self-adjoint operators (the
Friedrichs model) $h_{\mu_\alpha}(p),\,  p\in {\T}^3,$ acting in
$L_2(\T^3)$ by
\begin{equation}\label{h_alpha}
 h_{\mu_\alpha}(p)=h_\alpha^{0}(p)-\mu_\alpha v_\alpha,
 \end{equation}
 where
$$
 (h^0_\alpha(p)f)(q)=u_p^{(\alpha)}(q)f(q),\quad
 f\in L_2(\T^3),$$
and $v_{\alpha},\alpha=1,2,$ are non-local interaction operators
\begin{align*}
 \label{poten} (v_{\alpha}f)(q)=\varphi_\alpha(q)\int_{\T^3}
\varphi_\alpha(t)f(t)dt,\quad
 f\in L_2(\T^3).
 \end{align*}
\begin{remark} The spectrum and resonances of the
Friedrichs model are studied in \cite{ALzMarX06,Fri48,Fad64,Yaf92}.
\end{remark}

Let ${\C}$ be the field of complex numbers. Set
\begin{align*}
&m_\alpha(p)=\min_{q\in {\T}^3}u^{(\alpha)}_p(q),\quad
M_\alpha(p)=\max_{q\in {\T}^3} u^{(\alpha)}_p(q),\\
 &m=\min_{p,q\in
\T^3} u(p,q) ,\quad M=\max_{p,q}u(p,q).
\end{align*}
 and
\begin{equation*}\label{Lamb}
\Lambda_{\alpha}(p,z)=\int_{{\T}^3}
\frac{\varphi_\alpha^2(t)dt}{u_p^{(\alpha)}(t)-z},\quad
p\in\T^3,\,\, z \in \C\setminus [m_\alpha(p),M_\alpha(p)].
\end{equation*}

\begin{remark}
Note that by part $(i)$ of Hypothesis \ref{hypoth}  all third order
partial derivatives of the function $\Lambda_{\alpha}(\cdot,z)
,\,z<m, $ belong to $\cB (\theta,\T^3)$ and if $z=m$, then
$\Lambda_{\alpha}(\cdot,m)$ is continuous in $\T^3.$
\end{remark}

In order to prove the finiteness and infiniteness of eigenvalues
below the bottom of the essential spectrum of $H$  we assume the
following
\begin{hypothesis}\label{Lambda}
Assume that the function $\Lambda_\alpha(\cdot,m)$ has a unique
maximum at the origin such that for some $c>0$ the following
inequality holds
\begin{equation*}
\Lambda_\alpha(0,m)-\Lambda_\alpha(p,m)>c|p|^2,\quad 0\neq p\in
U_\delta(0).
\end{equation*}
\end{hypothesis}

 Recall (see, e.g., \cite{ReSiIV,ALkM}) that a complex-valued bounded
function $\varepsilon:\T^3\rightarrow \C$ is called conditionally
negative definite if $\varepsilon(p)=\overline{\varepsilon(-p)}$ and
\begin{equation*}\label{nn}
  \sum_{i,j=1}^{n}\varepsilon(p_i-p_j)z_i\bar z_j\le 0
 \end{equation*}
for any $n\in \N$, for all  $p_1, p_2, \dots, p_n\in \T^d$ and all
${\bf z}=(z_1, z_2,\dots, z_n)\in \C^n$ satisfying
$\sum_{i=1}^nz_i=0$.
\begin{remark}
Assume that $\varepsilon(\cdot)$ is a real-valued conditionally
negative definite  function on ${\T}^3$ having a unique
non-degenerate minimum at the origin and such that all third order
partial derivatives of $\varepsilon(\cdot)$ are continuous and
belong to $\cB (\theta,\T^3)$. Let the function $u(\cdot,\cdot)$ be
of the form
\begin{equation*}
u(p,q)=\varepsilon (p)+ \varepsilon (p-q)+ \varepsilon (q).
\end{equation*}
 Then Hypotheses  \ref{hypoth} and \ref{Lambda}
are fulfilled (see Lemma 5.3 in \cite{ALzMarX06}).
\end{remark}
\begin{definition}\label{resonance0}
Let part $(i)$ of Hypothesis \ref{hypoth} be fulfilled and let
$\varphi_\alpha \in \cB (\theta,\T^3),\,\,\frac{1}{2}<\theta\leq 1.$
 The operator $h_{\mu_\alpha} (0)$ is said to
have a threshold  energy resonance if the number  $1$ is an
eigenvalue of the operator
$$
(\mathrm{G}_\alpha\psi)(q)=\mu_\alpha \varphi_\alpha(q)
\int_{{\T}^3} \frac{\varphi_\alpha(t)\psi(t)dt}
{u_0^{(\alpha)}(t)-m},\quad \psi\in {C(\T^3)}
$$
and the associated eigenfunction $\psi $ (up to constant factor)
satisfies the condition $\psi(0)\neq 0.$
\end{definition}

\begin{remark}
Let part $(i)$ of Hypothesis \ref{hypoth} be fulfilled and let
$\varphi_\alpha \in \cB (\theta,\T^3),$
$\,\,\frac{1}{2}<\theta\leq 1.$\\
(i) If $\varphi_\alpha(0)\not=0$ and $\mu_\alpha=\mu_\alpha^0,$ then
the operator $h_{\mu_\alpha^0}(0)$ has a threshold energy resonance
and the function
\begin{equation}\label{f0f1}
f(q)=\frac{\varphi_\alpha(q)}{u_0^{(\alpha)}(q)-m},
\end{equation}
obeys the equation $ h_{\mu_\alpha^0}(0)f=mf$ and  $ f\in
L_1(\T^3)\setminus L_2(\T^3)$
(see Lemma \ref{delta=0}).\\
(ii) If $\varphi_\alpha(0)=0$ and $\mu_\alpha=\mu_\alpha^0,$ then
 the operator $h_{\mu_\alpha^0}(0)$ has a threshold eigenvalue
 and the function $f,$
defined by \eqref{f0f1}, obeys the equation $
h_{\mu_\alpha^0}(0)f=mf$ and $ f\in L_2(\T^3)$ (see Lemma
\ref{delta=0}).
\end{remark}

Let $ \tau_{ess}(H)$ be the bottom of the essential spectrum
 and $N(z)$ be the number of eigenvalues of $H$ lying below
 $z \leq \tau_{ess}(H).$

Set
\begin{equation*}\label{mu.alpha}
 \quad \mu_\alpha^0=\Lambda_\alpha^{-1}(0,m).
\end{equation*}

The main results of the present paper are as follows:
\begin{theorem}\label{fin}  Let Hypotheses \ref{hypoth} and
\ref{hyp.varphi} be fulfilled and $\mu_\alpha
=\mu_\alpha^0,\, \alpha=1,2.$\\
 (i)  Assume that Hypothesis \ref{Lambda} is fulfilled
  and $\varphi_1(0)\varphi_2(0)=0.$
 Then the
operator $H$ has a finite number of eigenvalues outside of the essential spectrum.\\
 (ii) Assume $\varphi_\alpha(0)\neq 0$ for any $\alpha=1,2$ and that
 Hypothesis \ref{Lambda} is fulfilled. Then the discrete
spectrum of $H$ is infinite and the function $N(\cdot)$  obeys the
relation
\begin{equation} \label{asym.K}
\lim\limits_{z \to m-0}\frac{N(z)}{|\log |m-z||}={\cU}_0 \quad
(0<{\cU}_0 <\infty).
\end{equation}
\end{theorem}
\begin{remark}
In fact in \cite{ALzMarX06} a result analogue, to part (i)  of
Theorem \ref{fin},  has been proven for the three-particle
Schr\"odinger operators on the lattice $\Z^3$ in the case, where the
function $u(\cdot,\cdot)$ is of the form
\begin{equation*}
u(p,q)=\varepsilon (p)+ \varepsilon (p-q)+ \varepsilon (q),
\end{equation*}
 and
\begin{align*}\label{of forms}
\varepsilon (q)=3-cos q_1-cos q_2 -cos q_3,\quad q=(q_1,q_2,q_3) \in
{\T}^3,
\end{align*} $\varphi_\alpha(\cdot)\equiv \const.$
\end{remark}
\begin{remark} The constant ${\cU}_0$ does not depend on the
functions $\varphi_\alpha$  and is given as a positive function
depending only on the ratios $\frac {l_\alpha}{l},\alpha=1,2$.
\end{remark}

\begin{remark} Clearly, the infinite cardinality of the
 discrete spectrum of $H$  lying on the l.h.s. of $m$
follows automatically from the positivity of  \, ${\cU}_{0}.$
\end{remark}
\begin{remark}
 Notice that under assumptions of Theorem \ref{fin}
 for all nonzero  $p\in \T^3$ the operator $h_{\mu_\alpha^0}(p)-mI$
  is strictly positive
and so that the operator $h_{\mu_\alpha^0}(0)$ corresponding to zero
value of $p$ is a unique operator whose spectrum attains  the
bottom of the essential spectrum of $H.$  Moreover  the essential
spectrum $\sigma_{ess}(H)$ of $H$  consists only of the
three-particle continuum
 $[m,M].$
\end{remark}

\section{Threshold analysis of the family of  Friedrichs models
 $h_{\mu_\alpha}(p)$}

In this section for the  reader's convenience   we recall some
results concerning  the families of  Friedrichs models
$h_{\mu_\alpha}(p),\,  p\in {\T}^3,$ defined by \eqref{h_alpha},
from the paper \cite{ALzMarX06}.

In accordance to Weyl's theorem
 the essential
  spectrum of the operator
$h_{\mu_\alpha}(p)$ fills the following interval on the real axis:
$$
\sigma_{ess}(h_{\mu_\alpha}(p))=[m_\alpha(p),M_\alpha(p)].
$$

 For any $p \in \T^3$ we
define an analytic  function $\Delta_{\mu_\alpha}(p,\cdot)$ (the
Fredholm determinant
 associated to the operator $h_{\mu_\alpha}(p)$)  on
 ${\C} { \setminus } [m_\alpha(p),M_\alpha(p)]$ by
\begin{equation*}\label{det}
\Delta_{\mu_\alpha}(p,z)=1-\mu_{\alpha} \Lambda_{\alpha}(p,z).
\end{equation*}

The following lemma  describes whether the bottom of the essential
spectrum of $h_{{\mu_\alpha^0}}(0)$ is  a threshold energy resonance
or a threshold eigenvalue.

\begin{lemma}\label{delta=0}
 Assume part $(i)$ of Hypothesis
\ref{hypoth} and $ \varphi_{\alpha} \in \mathcal{B}(\theta,\T^3).$\\
 (i) For any ${\mu_\alpha}>0$ and
$p\in \T^3$  the operator $h_{\mu_\alpha}(p)$ has an eigenvalue
$z
\in {\C} \setminus [m_\alpha(p),M_{\alpha}(p)]$  if and only if
$\Delta_{{\mu_\alpha}}(p,z)=0.$\\
(ii) the operator $h_{{\mu_\alpha}}(0)$ has a threshold energy
resonance (resp. threshold eigenvalue)  if and only if
  ${\mu_\alpha}=
{\mu_\alpha}^{0}$ and $\varphi_{\alpha}(0)\neq 0$ (resp.
 $\varphi_{\alpha}(0)=0$).
\end{lemma}

 The following
Lemma \ref{main0}  plays a crucial role in the proof of the
infiniteness (resp. finiteness) of the number of eigenvalues lying
below the bottom of the essential spectrum
 for a model operator $H$ associated
to a system of three-particles on the lattice $\Z^3$ interacting via
pair non-local potential.

\begin{lemma}\label{main0}
Assume that Hypotheses \ref{hypoth} and \ref{hyp.varphi} are
fulfilled.\\
 $(i)$ Let the operator $h_{\mu_\alpha^0}(0)$ have a threshold  energy
resonance. Then:\\
$(i_1)$  for all $p\in U_{\delta}(0) $ and $z \leq m$ the following
expansion holds
\begin{align*}
\Delta_{\mu^0_{\alpha}}(p,z)=\frac{4\sqrt{2}\pi^2 \mu_\alpha^0
\varphi_{\alpha}^2(0)}
 {l_\beta^{\frac{3}{2}} \det(U)^{\frac{1}{2}}}\sqrt{m_\alpha(p)-z}+
\Delta_{\mu^0_{\alpha}}^{res}(z)+
\Delta_{\mu^0_{\alpha}}^{res}(p,z),
\end{align*}
 where
$\Delta_\mu^{res}(m_\alpha(p)-z)=O((m_\alpha(p)-z)
^{\frac{1+\theta}{2}})$
as $m_\alpha(p)-z \to 0,$ $z<m_\alpha(p),$ and
 $\Delta_{\mu_{\alpha}}^{res}(p,z)=O(p^2)$ as $p\to 0$ uniformly in
$z\leq m_\alpha(p);$\\
$(i_2)$ for some $c_1,c_2>0$  the inequalities
\begin{equation*}\label{c<(.,.)<c}
c_1 |p| \leq \Delta_{\mu^0_{\alpha}}(p,m) \leq c_2 |p|, \quad p\in
U_\delta(0),
 \end{equation*}
\begin{equation*}\label{(.,.)>c}
\Delta_{\mu^0_{\alpha}}(p,m) \geq c , \quad p\in \T^3\setminus
U_\delta(0)
 \end{equation*}
hold.

(ii) Let $z=m$ be an  eigenvalue of $h_{\mu_\alpha^0}(0)$. Then:\\
$(ii_1)$ for any $p\in U_{\delta}(0) $ and $z \leq m$ the following
expansion holds
\begin{align*} \Delta_{\mu^0_{\alpha}}(p,z)=
\Delta_{\mu^0_{\alpha}}^{res}(z)+
\Delta_{\mu^0_{\alpha}}^{res}(p,z);
\end{align*}
\\
$(ii_2)$ the inequality
\begin{align*}
\Delta_{\mu^0_{\alpha}}(p,m)\geq c p^2,\quad p\in U_\delta(0),
\end{align*}
holds,  for some $c>0$.
\end{lemma}
\begin{remark}
Lemma \ref{main0} gives threshold energy expansions for the Fredholm
determinant, leading to different behaviors for a threshold energy
resonance resp. eigenvalue.
\end{remark}

\section{The essential  spectrum of the  operator $H$}

In this section we exhibit the location of the essential spectrum of
the model operator $H.$

Set
\begin{equation*}
  \Sigma=\cup_{\alpha=1}^2 \cup_{p\in
{\T}^3}\sigma_d(h_{\mu_\alpha}(p)) \cup [m,M],
\end{equation*}
 where $\sigma_d(h_{\mu_\alpha}(p))$ is the discrete spectrum of the
operator
    $h_{\mu_\alpha}(p),\,p\in \T^3.$

Let
 $\Phi_\alpha:L_ 2(({\T}^3)^2)\to
L_ 2({\T}^3),\,\alpha=1,2, $  be the operator defined by
\begin{align}\label{izom}
(\Phi_1 f)(q)= \int_{{\T}^3}\varphi_1(t) f(t,q)dt,\quad (\Phi_2
f)(p)=\int_{{\T}^3}\varphi_2(t) f(p,t)dt,\,\, f\in
L_2((\T^3)^2)\nonumber
\end{align}
 and
 denote by $\Phi^*_\alpha$ its adjoint.
Let $D_\alpha(z) $  be the multiplication operator by the function
$\Delta_{\mu_\alpha}(q,z)$ on $L_2(\T^3).$

 It is easy to prove the  equality
\begin{equation}\label{W}
I-\mu_\alpha \Phi_\alpha R_0(z)\Phi^*_\alpha=D_\alpha(z),
\quad z\in
\C\setminus [m,M],
\end{equation}
where $R_0(z)=(H_0-z{\bf I})^{-1}$ is the resolvent of
 $H_0$ and  $I$ resp. ${\bf I}$ is the identity operator
 on $L_2(\T^3)$ resp. $L_2((\T^3)^2).$

By Lemma \ref{delta=0} for any $z\in \C\setminus {\Sigma}$ the
inequality $\Delta_{\mu_\alpha}(p,z)\neq 0$ holds. Then  the
operator $D_{\alpha}(z),$ $z\in \C\setminus {\Sigma},$ is
invertible. Let $D^{-1}_{\alpha}(z)$  be its inverse.

Let $ \mathrm{T}( z),\, z \in \C\setminus \Sigma,$ act in
$L^{(2)}_2({\T}^3)$ with the entries
\begin{align*}
 &\mathrm{T}_{\alpha\alpha } (  z)=0, \quad \mathrm{T}_{\alpha\beta}(z)=
  \sqrt {\mu_\alpha\mu_\beta} D^{-1}_\alpha(z)\Phi_\alpha
  R_0(z) \Phi^*_\beta.
\end{align*}
\begin{lemma}\label{tz} For any $z \in \C\setminus {\Sigma}$
the operator $\mathrm{T}(z)$ is an Hilbert-Schmidt operator.
\end{lemma}
\begin{proof}
According to the fact that $\Phi_\alpha  \Phi^*_\beta,$ $\alpha\neq
\beta,$ is a compact integral operator one  checks that  for any $z
\in \C\setminus {\Sigma}$
 the operator $\Phi_\alpha
R_0(z)\Phi^{*}_\beta $ belongs to the Hilbert-Schmidt class
$\Sigma_2$. Since the operator $D^{-1}_\alpha(z),$  $z \in
\C\setminus {\Sigma},$ is bounded, the operator
$\mathrm{\mathrm{T}}_{\alpha\beta}(z)$ also belongs to $\Sigma_2.$
\end{proof}

The following theorem describes the essential spectrum of the
operator $H$ by the spectrum of the family
$h_{\mu_\alpha}(p),\,p\in\T^3,\,\alpha=1,2.$
\begin{theorem}\label{ess1}
For the essential spectrum $\sigma_{ess}(H)$  of the operator $H$
the following equality holds
 $$
\sigma_{ess}(H)=\cup_{\alpha=1}^2 \cup_{p\in
{\T}^3}\sigma_d(h_{\mu_\alpha}(p)) \cup [m,M].
$$
\end{theorem}
\begin{proof}  The proof of theorem consists of two steps.
The inclusion $\Sigma  \subset \sigma_{ess}(H)$ is proven using
Weyl's criterion, as given in \cite{ALzMmn05} (we omit the details).
Let us prove the  inclusion $ \sigma_{ess}(H) \subset \Sigma $.

Denote by $R(z)=(H-z{\bf I})^{-1}$ the resolvent of the operator
$H.$
  The  well known   resolvent equation has the form
\begin{equation}\label{Rez}
R(z)=R_0(z)+R_0(z)(\mu_1 V_1+\mu_2 V_2)R(z).
\end{equation}

We observe that $V_\alpha=\Phi^*_\alpha\Phi_\alpha$.
 Multiplying  \eqref{Rez} from the left side by
 $\sqrt{\mu_\alpha}\Phi_\alpha $ and setting
  $\cR_{\alpha}(z) \equiv \sqrt{\mu_\alpha}\Phi_\alpha R(z)$
 we get the system of equations
\begin{equation}\no
\cR_{\alpha}(z)=\sqrt{\mu_\alpha}\Phi_\alpha R_0(z)+
\sqrt{\mu_\alpha}\Phi_\alpha R_0(z)(\sqrt{\mu_1 }\Phi^{*}_1
\cR_1(z)+\sqrt{\mu_2}\Phi^{*}_2 \cR_2(z)),\, \alpha=1,2,
\end{equation}
or the following system of three equations
\begin{align}\label{Rez1}
&(I-\mu_\alpha\Phi_{\alpha}R_0(z)\Phi^{*}_{\alpha})\cR_{\alpha}(z)=
\sqrt{\mu_\alpha}\Phi_{\alpha}R_0(z)+
\sqrt{\mu_\alpha\mu_\beta}\Phi_{\alpha}R_0(z) \Phi^{*}_{\beta}
\cR_{\beta}(z), \, \alpha=1,2.
\end{align}

As we mentioned above \eqref{W} the operator
$(I-\mu_\alpha\Phi_{\alpha}R_0(z)\Phi^{*}_{\alpha})\equiv
D_\alpha(z),$ $z\in \C\setminus {\Sigma},$ is a multiplication
operator on $L_2(\T^3)$ and is invertible.

 Multiplying the equality \eqref{Rez1} from the left by
 $D^{-1}_\alpha(z)$
  we get the Faddeev type
equation
\begin{equation}\label{Rez2}
{\cR}(z)={\cR}_0(z)+\mathrm{T}(z){\cR}(z),
\end{equation}
where  ${\cR}(z)=(\cR_{1}(z),\cR_{2}(z))$ and ${\cR}_0(z)=(
\sqrt{\mu_1}D^{-1}_1(z) \Phi_{1}R_0(z),\sqrt{\mu_2} D^{-1}_2(z)
\Phi_{2}R_0(z))$ are vector operators.

 From \eqref{Rez} we have the following representation for the
 resolvent
\begin{equation} \label{Rez4}
R(z)=R_0(z)+R_0(z)(\sqrt{\mu_1 }\Phi^{*}_1
\cR_1(z)+\sqrt{\mu_2}\Phi^{*}_2 \cR_2(z)).
\end{equation}

 Let ${\cI}$ be the  identity operator in $L^{(2)}_2(\T^3).$
Since $||\mathrm{T}(z)|| \to 0$ as $z\to \infty$ the operator
$\mathrm{T}(z)$ is a compact operator-valued function on $
\C\setminus {\Sigma}$ and ${\cI }-\mathrm{T}(z)$ is invertible if
$z$ is real and either very negative or very positive. The analytic
Fredholm theorem (see, e.g., Theorem $VI.14$ in \cite{ReSiIV})
implies that there is a discrete set $S \subset \C\setminus {\Sigma}
$ so that $({\cI }-\mathrm{T}(z))^{-1}$ exists and is analytic in
$\C\setminus ( \Sigma \cup S)$ and meromorphic in $\C\setminus
{\Sigma}$ with finite rank residues. Thus the function
$({\cI}-\mathrm{T}(z))^{-1} {\cR}_0(z)\equiv F(z)$ is analytic in
$\C\setminus ( \Sigma \cup S)$ with finite rank residues at the
points of $S$.

Let $z \notin S$, $\Img z\neq 0$, then by \eqref{Rez2},
\eqref{Rez4} we have $F(z)={\cR}(z).$ In particular,
$$
R(z)(H-z{\bf
I})=(R_0(z)+R_0(z)\sum_{\alpha=1}^{2}\sqrt{\mu_{\alpha}}
\Phi^*_{\alpha}\cR_\alpha(z) ) (H-z{\bf I})={ \bf I}.
$$

 By analytic
continuation, this holds for any $z \notin {\Sigma}\, \cup S. $ We
conclude that, for any such $z,$ the operator $H-z{\bf I}$ has a
bounded inverse. Therefore  $\sigma(H) \setminus  \Sigma $ consists
of
 isolated points and  only the  frontier points of $\Sigma$
can  possibly  be  limit points. Finally, since $R(z)$ has finite
rank residues at  $z \in S$, we conclude that $\sigma(H) \setminus
\Sigma $ belongs to the discrete spectrum $\sigma_{d}(H)$ of $H,$
which completes the proof.
\end{proof}
\begin{corollary}\label{ess2}
 Let the assumptions of Theorem \ref{fin} be fulfilled.
Then  the essential spectrum $\sigma_{ess}(H)$ of $H$ consists of
the interval  $[m,M].$
\end{corollary}
    \begin{proof}
 Hypothesis \ref{Lambda} and the equality $\mu_\alpha^0=\Lambda_\alpha^{-1}(0,m)$
yield $\Delta_{\mu_\alpha^0}(0,m)=0$ and hence the inequality
$$
\Delta_{\mu^0_\alpha}(p,m)=\mu_\alpha^0\big
(\Lambda_\alpha(0,m)-\Lambda_\alpha(p,m)\big )>0,\quad 0\neq p\in
\T^3.
$$

Since the function $\Delta_{\mu^0_\alpha}(p,\cdot)$ is decreasing in
$(-\infty,m_\alpha(p))$  by Lemma \ref{delta=0} we obtain
 $h_{\mu_\alpha^0}(p)>m,\,0\neq p\in \T^3$.
This argument, together with Theorem \ref{ess1}, completes  the
proof of Corollary \ref{ess2}.
\end{proof}

\subsection{Birman-Schwinger principle}

We recall that $ \tau_{ess}(H)$ denotes the bottom of the essential
spectrum
 and $N(z)$ the number of eigenvalues of $H$ lying below
 $z \leq \tau_{ess}(H).$

For a bounded self-adjoint operator $B,$ we define $n(\lambda,B)$ by
$$
n(\lambda,B)=sup\{ dim F: (Bu,u) > \lambda,\, u\in F,\,||u||=1\}.
$$
$n(\lambda,B)$ is equal to  infinity if $\lambda$ is in the
essential spectrum of $B$ and if $n(\lambda,B)$ is finite, it is
equal to the number of the eigenvalues of $B$ larger than $\lambda$.
 By the definition of $N(z)$ we have $$
N(z)=n(-z,-H),\,-z > -\tau_{ess}(H). $$

In our analysis of the spectrum of $H$ the crucial role is played by
the self-adjoint compact operator  $ T( z),\, z <\tau_{ess}(H)$ in
$L^{(2)}_2({\T}^3)$ with the entries
\begin{align*}
 &T_{\alpha\alpha } (  z)=0, \quad T_{\alpha\beta}(z)=
  \sqrt {\mu_1\mu_2} D^{-\frac{1}{2}}_\alpha(z) \Phi_\alpha
  R_0(z) \Phi^*_\beta D^{-\frac{1}{2}}_\beta(z).
\end{align*}

The following lemma is a realization of the well known
Birman-Schwinger principle for the operator $H$ (see
\cite{ALzM04,Sob93,Tam94} ).
\begin{lemma}\label{b-s}
The operator $T(z)$ is compact and continuous in $z<\tau_{ess}(H)$
and
$$ N(z)=n(1,T(z)). $$
\end{lemma}
\begin{proof}
This lemma is deduced by the same arguments as well as in
\cite{ALzM04,Sob93}. Set $ V=\mu_{1}V_1+\mu_{2}V_2.$ Since for any
$z<\tau_{ess}(H)$ the following relation
\begin{align*}
 &f\in L_2((\T^3)^2),\,(Hf,f)<z(f,f) \Leftrightarrow
 (R^{\frac{1}{2}}_0(z)VR^{\frac{1}{2}}_0(z)g,g)>(g,g),\quad
\\&g=(R_0(z))^{-\frac{1}{2}}f,  \,g\in
L_2((\T^3)^2),\,\,
\end{align*}
holds, the quantity $N(z)$ in
 question coincides with
$n(1,R^{\frac{1}{2}}_0(z)VR^{\frac{1}{2}}_0(z)),$ that is,
\begin{equation*}\label{tenglik}
N(z)=n(1,R^{\frac{1}{2}}_0(z)VR^{\frac{1}{2}}_0(z)).
\end{equation*}

Decompose
$
R^{\frac{1}{2}}_0(z)VR^{\frac{1}{2}}_0(z))=B^{*}B, $
 with the vector operator $B:L_2((\T^3)^2)\to L^{(2)}_2(\T^3)$
defined by
\begin{equation*}
B=\big (\sqrt{\mu_{1}}\Phi_{1}R^{\frac{1}{2}}_0(z),
\,\sqrt{\mu_{2}}\Phi_{2}R^{\frac{1}{2}}_0(z) \big).
\end{equation*}

One can see that the operator $M(z)=BB^*$
 acts in $L^{(2)}_2(\T^3)$ with
the entries
$$ M_{{\alpha \beta}}(z)
=\sqrt{\mu_{\alpha}\mu_{\beta}}\Phi_\alpha
R_0(z)\Phi^{*}_\beta,\quad \alpha,\beta =1,2. $$

Both operators $B^{*}B$ and $M(z)=BB^{*}$ have the same nonzero
eigenvalues with the same multiplicities.
 We use this
argument to obtain the equality
\begin{equation}\label{N=M}
N(z)=n(1,M(z)).
\end{equation}
We decompose $M(z)$ into the sum $M(z)=M_0(z)+K(z),$ where
\begin{equation*}
 M_0(z)=diag \{ M_{{11}}(z),M_{{22}}(z) \},\quad
K(z)=\left (  \begin{array}{ll}
 0 \quad  M_{12}(z)\\
M_{21}(z) \,\, 0
\end{array}\right ).
\end{equation*}

 By \eqref{W} the operator
  $ I-M_{{\alpha\alpha}}(z)=D_\alpha(z),$ $z<\tau_{ess}(H),$ is invertible
and by $\Delta_{\mu^{0}_{\alpha}}(p,z)>0,\, p\in
\T^3,\,z<\tau_{ess}(H),$ it is positive and a direct calculation
shows that $n(1,M(z))=n(1,({\cI} -M_0(z))^{-\frac{1}{2}}K(z)({\cI}
-M_0(z))^{-\frac{1}{2}}).$
 Then, to finish the proof, it suffices to
coincide the equality $T(z)=({\cI} -M_0(z))^{-\frac{1}{2}}K(z)({\cI}
-M_0(z))^{-\frac{1}{2}}$  and \eqref{N=M}.
\end{proof}

\section{ The finiteness of the  number of eigenvalues of the  operator $H$.}
In this section we will prove  part (i) of Theorem  \ref{fin} (the
finiteness of the number of eigenvalues ).
 We starts the proof  with the following assertion
\begin{theorem}\label{upper.eige}
 The
operator $H$ has no eigenvalues lying on the right hand side of the
essential spectrum $\sigma_{ess}(H).$
\end{theorem}
\begin{proof}
Since $V=\mu_1V_1+\mu_2V_2$ is a positive operator and $\sup
(\sigma_{ess}(H))=\sup (\sigma(H_{0}))=M$  we have that the operator
$H=H_0-V$ has no eigenvalues larger than $M$.
\end{proof}

Now  we  prove that  $H$ has a finite number of eigenvalues on the
left hand side of its essential spectrum.

To do this, we use the following two lemmas.
\begin{lemma}\label{U.ineq} Let Hypothesis \ref{hypoth} be
fulfilled. Then there exist numbers $C_1, C_2,C_3>0$ and $\delta>0$
such that the following inequalities hold
\begin{align*}
&(i)\quad C_1 (|p|^2+|q|^2)\leq u(p,q)-m \leq C_2 (|p|^2+|q|^2)\quad
\mbox{for all} \quad p, q\in U_\delta(0),\\
 &(ii)\quad u(p,q)-m \geq C_3\quad \mbox{for all} \quad(p, q)\notin
U_\delta(0)\times U_\delta(0).
\end{align*}
\end{lemma}
\begin{proof}
 By Hypothesis \ref{hypoth} the point $(0,0)\in (\T^3)^2$ is
a unique non-degenerated minimum point of $u.$ Then  there exist
positive  numbers $C_1,C_2,C_3$ and a $\delta-$neighborhood of
$(p,q)=(0,0)\in (\T^3)^2$ so that $(i)$ and $(ii)$ hold true.
\end{proof}

\begin {lemma}\label{G-S}  Let the conditions in part $(i)$ of
Theorem \ref{fin} be fulfilled.
 Then  the operator $T(z)$
  belongs to the Hilbert-Schmidt class and
is continuous from the left up to $z=m$.
\end{lemma}
\begin{proof}
We prove  Lemma \ref{G-S}
 in the case $\mu_\alpha =\mu_\alpha^0,$
  and $\varphi_1(0)=0$, $\varphi_2(0)\neq
 0$ (the other cases are handled in a similar way).

Since the function $\varphi_1\in C^{(2)}(\T^3)$ is either even or
odd and $\varphi_1(0)=0$ we have $|\varphi_1(p)|\leq C|p|$ for any
$p\in \T^3$ and for some $C>0$.
 By virtue of  Lemmas \ref{main0} and \ref{U.ineq}
 the kernel of the operator $T_{12}(z),\,z\leq m,$ is
estimated by
$$ C\big (
 \frac{\chi_{\delta}(p)}{|p|}+1\big )\big (
\frac{|q| \chi_{\delta}(p)\chi_{\delta}(q)}{p^2+q^2}+1 \big )\big
(\frac{\chi_{\delta}(q)}{|q|+1}
 \big),
$$ where  $ \chi_{\delta}(p)$ is the characteristic function of
$U_\delta(0).$

Since the
 function is square-integrable on $(\T^3)^2$  we
have that $T_{12}(z)$ and $T_{21}(z)=T^*_{12}(z)$ are
Hilbert-Schmidt operator.

The kernel function of $T_{\alpha\beta}(z)$ is continuous in $p,q
\in \T^3$ and in $z<m$ and is square-integrable
 on $(\T^3)^2$ as $z\leq m$.
Now the continuity of the operator $T_{\alpha\beta}(z)$ from the
left up to $z=m$   follow using  Lebesgue's dominated convergence
theorem.
\end{proof}

We are now ready for the

 {\bf Proof of $(i)$ of Theorem \ref{fin}.}
Let the conditions in part $(i)$ of Theorem \ref{fin} be fulfilled.
   By Lemma \ref{b-s}  we have
$$
N(z)=n(1,T(z))\,\,\mbox{as}\,\,z<m
$$
and by Lemma \ref{G-S} for any $\gamma\in [0,1)$ the number
$n(1-\gamma,T(m)) $ is finite. Then for all $z<m$ and $\gamma \in
(0,1)$ we have $$ N(z)=n(1,T(z))\leq
n(1-\gamma,T(m))+n(\gamma,T(z)-T(m)). $$

This relation can easily be obtained by using   Weyl's inequality
\begin{equation}\label{Weyl}
n(\lambda_1+\lambda_2,A_1+A_2)\leq n(\lambda_1,A_1)+n(\lambda_2,A_2)
\end{equation}
for the sum of compact operators $A_1$ and $A_2$ and for  positive
numbers $\lambda_1$ and $\lambda_2.$

By Lemma \ref{G-S} the operator $T(z)$ is continuous from the left
up to $z=m$ and hence
$$ \lim_{z\to m-0} N(z)= N(m)\leq n(1-\gamma,T(m))\,\, \mbox{for
all}\,\, \gamma \in (0,1). $$

Thus
 $$N(m)\leq n(1-\gamma,T(m))<\infty.$$ The latter
inequality and Theorem \ref{upper.eige} prove the assertion $(i)$ of
Theorem \ref{fin}. \qed

\section{Asymptotics for the number of
eigenvalues of the  operator $H$.}

In this section we shall derive the   asymptotics \eqref{asym.K} for
the number of eigenvalues of $H$.

 By Hypothesis \ref{hypoth} we get
\begin{equation}\label{asymp1}
 u(p,q)=m+\frac{1}{2}\big(
l_1(Up,p)+2l(Up,q)+l_2(Uq,q)\big )+O(|p|^{3+\theta}+|q|^{3+\theta})
\end{equation}
as  $p,q\rightarrow 0$ and
$$m_{\alpha}(p)=m+\frac{l_1
l_2-l^2}{2l_\beta}(Up,p)+O(|p|^{3+\theta}) \quad \mbox{as}\quad p
\to 0 .$$
 Applying the latter asymptotics for $m_{\alpha}(p)$ and using Lemma \ref{main0}
  we have
\begin{equation}\label{asymp2}
\Delta_{\mu^0_\alpha}(p,z) = \frac{4\pi^2 \mu^0_\alpha
\varphi^2_\alpha(0)}
 {l_\beta^{{3}/{2}} \det(U)^{\frac{1}{2}}}
 \left [ n_\alpha (Up,p) -2(m-z) \right ]^{\frac{1}{2}}+
 O\big ((|p|^2+|m-z|)^{\frac{1+\theta}{2}} \big)
 \end{equation}
as $p,|m-z|\rightarrow 0,$ where
\begin{equation*}
n_\alpha={(l_1l_2-l^2)}/{l_\beta}.
\end{equation*}

 Let $T(\delta;|m-z|)$ be the integral operator in $L_2^{(2)}({\T}^3)$  with the kernel
\begin{align*}
&T_{\alpha\alpha}(\delta,|m-z|;p,q)=0,\\ &T
_{\alpha\beta}(\delta,|m\!\!-\!\!z|;p,q)\!\!=\!\!\mathrm{d}_0 \frac{
\hat \chi_\delta (p) \hat \chi_\delta (q) (n_\alpha(Up,p)+
2|m-z|)^{^{^{-\!\frac{1}{4}}}} (n_\beta  (Uq,q)+ 2|m-z|)^
{^{^{-\!\frac{1}{4}}}} } {l_\alpha(Up,p)+ 2l(Up,q)+l_\beta (Uq,q)
 +2|m-z|},
\end{align*}
where $\hat \chi_\delta(\cdot)$ is the  characteristic function of
the region $\hat U_\delta(0)=\{ p\in \T^3:\,\,
|U^{\frac{1}{2}}p|<\delta \}$ and
 $$ \mathrm{d}_0= \frac{{\det U}^{\frac{1}{2}}} {2 \pi^2}l_1^{\frac{3}{4}}
l_2^{\frac{3}{4}}.
 $$

\begin{lemma}\label{H-SH} Let the conditions in part $(ii)$ of Theorem \ref{fin} be
fulfilled. The operator $ T(z)-T(\delta; |m-z|)$ belongs to the
Hilbert-Schmidt class and is continuous in $z\leq m.$
\end{lemma}
\begin{proof}
Applying   asymptotics \eqref{asymp1}, \eqref{asymp2} and Lemmas
\ref{main0}, \ref{U.ineq} one can estimate the kernel of the
operator $ T_{\alpha\beta} (z)-T_{\alpha \beta}(\delta;
|m-z|),\,z\le m$ by the square-integrable function
\begin{equation*}
 C \Big (\frac{|p|^{1+\theta}+|q|^{1+\theta}}{|p|^{\frac{1}{2}}(p^2+q^2)|q|^{\frac{1}{2}}}+
\frac{|m-z|^{\frac{\theta}{2}}(p^2+q^2)^{-1}}{(|p|^2+|m-z|)^{\frac{1}{4}}
(|q|^2+|m-z|)^{\frac{1}{4}}}+1 \Big).
\end{equation*} Hence
the operator $ T_{\alpha\beta} (z)-T_{\alpha \beta}(\delta; |m-z|)$
 belongs to the Hilbert-Schmidt class for all
$z \leq m.$ In combination with the continuity of the kernel of the
operator in  $z<m$ this  gives   the continuity of $
T(z)-T(\delta;|m-z|)$ in  $z\leq m.$
\end{proof}

Let us now recall some results from \cite{Sob93}, which are
important in our work.

Set  ${\bf \sigma}=L_2(\S^2),$ where $
\S^2$ being unit sphere in $\R^3,$ and ${\bf \sigma}^{(2)}={\bf
    \sigma}\oplus {\bf \sigma}.$

Let ${\bf S}_{{\bf r}},\,\mathbf{r}>0,$  be the integral operator in
$L_2((0,{\bf r}), {\bf \sigma}^{(2)})$ with the kernel
$S_{\alpha\beta}(y,t),\,y=x-x',\,x,x'\in (0,{\bf r}),\,\,t=<\xi,
\eta>,\,\xi, \eta \in \S^2,$ where
\begin{equation}\label{sobolev}
 S_{\alpha\alpha}(y,t)=0,\quad
 S_{\alpha\beta}(y,t)=(2\pi)^{-2}\frac{u_{\alpha\beta}}
{\cosh(y+r_{\alpha \beta})+s_{\alpha\beta}t},
\end{equation}
\begin{align*}& u_{\alpha\beta}=u_{\beta\alpha}=\big(
\frac{l_1l_2}{l_1l_2-l^2} \big)^{\frac{1}{2}},\,
 r_{\alpha\beta}=\frac{1}{2} \log
\frac{l_\alpha}{l_\beta} ,\,
s_{\alpha\beta}=s_{\beta\alpha}=\frac{l} {\sqrt{l_1l_2}},\\
&\alpha,\beta=1,2.
\end{align*}
Let $\hat {\bf S}(\lambda),\,\lambda\in \R,$ be the integral
operator on ${\bf \sigma}^{(2)}$   whose kernel depends on the
scalar product $t=<\xi,\eta>$ of the arguments $\xi,\eta\in\S^2$ and
has the form
\begin{equation*}
\hat S_{\alpha\alpha}(\lambda;t)=0,\quad \hat S_{\alpha\beta}(
\lambda;t)=(2\pi)^{-2}\frac{u_{\alpha\beta} e^{ir_{\alpha\beta}\lambda}
\sinh[\lambda(arccos s_{\alpha\beta}t)]}
{\sqrt{1-s_{\alpha\beta}^2t}\sinh(\pi x)}.
\end{equation*}

 For $\mu>0,$ define
\begin{equation*}
 {U}(\mu)= (4\pi)^{-1}
\int\limits_{-\infty}^{+\infty} n(\mu,\hat{\bf S}(\lambda))d\lambda.
\end{equation*}
 This function was studied in detail \cite{Sob93} and is very
important for the proof of the existence of the Efimov effect. In
particular, it was proved  that $U(\cdot)$ is
continuous, $U(1)>0$ if $u_{12}>1$ and
$$ \lim\limits_{{\bf r}\to
\infty} \frac{1}{2}{\bf r}^{-1}n(\mu,{\bf S}_{\bf r})={U}(\mu).$$

Part $(ii)$ of Theorem \ref{fin} will be deduced by a perturbation
argument based on the following lemma (see Lemma 4.7 in
\cite{Sob93}). For completeness, we reproduce the proof given there.
 \begin{lemma}\label{comp.pert}
 Let $A (z)=A_0 (z)+A_1 (z),$ where $A_0(z)$ resp. $A_1(z)$ is
compact and continuous in $z<m$ resp. $z\leq m.$  Assume that for
some function $f(\cdot),\,\, f(z)\to 0,\,\, z\to m-0$ the limit
$$ \lim_{z\rightarrow m-0}f(z)n(\lambda,A_0 (z))=l(\lambda), $$
 exists and $l(\lambda)$ is continuous in $\lambda>0.$ Then the same limit
exists for $A(z)$ and $$ \lim_{z\rightarrow m-0}f(z)n(\lambda,A
(z))=l(\lambda).
$$
\end{lemma}
\begin{proof}
According to Weyl's inequality \eqref{Weyl} for any $\varepsilon\in (0,1)$ we have
\begin{equation*}
n(\lambda,A(z))\leq n((1-\varepsilon)
\lambda,A_0(z))+n(\varepsilon\lambda,A_1(z)),
\end{equation*}
 \begin{equation*}
  n(\lambda,A(z))\geq n((1+\varepsilon)\lambda,A_0(z))
  -n(\varepsilon\lambda,A_1(z))
\end{equation*}
Since the operator $A_1(z)$ is compact and continuous in  $z\leq m$,
we obtain
 \begin{equation*}
  l((1+\varepsilon)\lambda)\leq \lim_{z\to m-0}\inf f(z)
  n(\lambda,A(z))\leq \lim_{z\to m-0}\sup f(z) n(\lambda,A(z))
\leq l((1-\varepsilon)).
\end{equation*}
Then  the continuity of the function $l(\lambda)$ in $\lambda>0$
completes the proof of Lemma \ref{comp.pert}.
\end{proof}

\begin{remark}\label{comp.pert0} Since $\cU(\cdot)$ is continuous in $\mu,$ according
to Lemma \ref{comp.pert}  any perturbations of the operator $A_0(z)$
defined in Lemma \ref{comp.pert}, which is compact and continuous up
to $z=m$ do not contribute to  asymptotics \eqref{asym.K}. During the
proof of Theorem \ref{main} we use this fact without further
comments.
\end{remark}

 The following theorem is basic for the proof of the asymptotics
\eqref{asym.K}.
\begin{theorem}\label{main} Let the conditions in part $(i)$ of Theorem \ref{fin}
be fulfilled. Then the equality
$$ \lim\limits_{|m-z|\to 0}
|log|z-m||^{-1}n(\mu,T(\delta; |m-z|)) ={U}(\mu),\quad \mu>0, $$
holds.
\end{theorem}
\begin{proof}
 The space of vector-functions
$w=(w_1,w_2)$ with coordinates having support in $\hat U_\delta(0)$
 is an invariant subspace
for the operator $T(\delta|m-z|).$

Let  $\hat T_0(\delta;|m-z|)$ be the restriction of the integral
operator $T (\delta|m-z|)$ to the subspace $L^{(2)}_2(\hat
U_{\delta}(0)).$ One verifies  that the operator $\hat
T_0(\delta;|m-z|)$ is unitarily equivalent to the integral operator
$T_0(\delta;|m-z|)$ in $L^{(2)}_2(\hat U_{\delta}(0))$ with the
kernel
\begin{align*}
&T^{(0)}_{\alpha\alpha}(\delta,|m-z|;p,q)=0,\\
&T^{(0)} _{\alpha\beta}(\delta,|m-z|;p,q)= \mathrm{d}_1
 \frac{ (n_\alpha p^2+ 2|m-z|)^{-1/4}
(n_\beta q^2+2|m-z|)^ {-1/4} } {l_\alpha p^2+ 2l(p,q)+l_\beta q^2
+2|m-z|} ,
\end{align*}
where
 $$ \mathrm{d}_1= ({2 \pi^2})^{-1}{l_1^{{3}/{4}}
l_2^{{3}/{4}}} .
 $$

Here the equivalence is performed by the unitary dilation $${\bf
Y}=diag\{Y_1,Y_2,\}:L^{(2)}_2( U_{\delta}(0))
 \to L^{(2)}_2(\hat U_{\delta}(0)),\quad
 (Y_\alpha f)(p)=f(U^{-\frac{1}{2}}p).
 $$

The operator $T_0(\delta;|m-z|)$ is unitary equivalent
 to the integral operator $T_1(r)$ acting in $L_2^{(2)}(U_r(0)),$ where
 $r=|m-z|^{-\frac{1}{2}}$ and
$U_r(0)=\{ p\in \R^3:|p|<r\},$  with the kernel
\begin{align*}
&T^{(1)}_{\alpha\alpha}(r;p,q)=0,\\
&T^{(1)}_{\alpha\beta}(r;p,q)= \mathrm{d}_1 \frac{  (n_\alpha
p^2+2)^{-1/4} (n_\beta  q^2+2)^ {-1/4} } {l_\alpha p^2+
2l(p,q)+l_\beta q^2 +2} .
\end{align*}

The equivalence is performed by the unitary dilation
 $${\bf B}_r=diag\{B_r,B_r\}:L_2^{(2)}
(U_\delta(0)) \to L_2^{(2)}(U_r(0)),\quad
 (B_r f)(p)=(\frac{r}{\delta})^{-3/2}f(\frac{\delta}
{r}p).$$

Further, we may replace
 $$(n_\alpha p^2+2)^{-1/4},\,
(n_\beta q^2+2)^{-1/4} \quad \mbox{ and}\quad l_\alpha p^2+
2l(p,q)+l_\beta q^2
 +2$$
 by
$$(n_\alpha p^2)^{-1/4}(1-\chi_1(p)),\,\, (n_\beta
q^2)^{-1/4}(1-\chi_1(q))
 \quad \mbox{ and}\quad
l_\alpha p^2+ 2l(p,q)+l_\beta q^2 ,$$
  respectively, since the error will be
a Hilbert-Schmidt operator  continuous up to  $z=m,$ where
$\chi_1(\cdot)$ is a characteristic function of the ball $U_1(0).$
Then we get the integral operator $T_2(r)$ in $L_2^{(2)}(U_r(0)
\setminus
U_1(0))$ with the kernel
\begin{align*}
&T^{(2)}_{\alpha\alpha}(r;p,q)=0,\\
& T^{(2)}_{\alpha\beta}(r;p,q)= \frac{\mathrm{d}_1}{(n_1
n_2)^{\frac{1}{4}}}\frac{ |p|^{-1/2}| q|^{-1/2}} {l_\alpha p^2+
2l(p,q)+l_\beta  q^2}.
\end{align*}

By the dilation
$${\bf M}=diag\{M,M \}:L_2^{(2)}(U_r(0) \setminus
U_1(0)) \rightarrow L_2((0,{\bf r}), {\bf \sigma}^{(2)}),\,{\bf
r}=1/2\big |log|m-z|\big |,
$$
 where
$(M\,f)(x,w)=e^{3x/2}f(e^{ x}w),\quad  x\in (0,{\bf r}),\, w \in {\S}^2,$ one sees that the
operator $T_2(r)$ is unitarily equivalent to the integral operator
${\bf S}_{{\bf r}}$ with entries \eqref{sobolev}.

 The difference of the operators ${\bf S}_{{\bf r}}$ and
$T(\delta; |m-z|)$ is compact (up to unitarily equivalence) and
hence, taking into account that ${\bf r}=1/2 | \log |m-z|$, we
obtain
$$ \lim\limits_{|z-m|\to 0} |log|z-m||^{-1}n(\mu,T(\delta; |z-m|))
={U}(\mu),\quad \mu>0 .$$
Theorem \ref{main} is proved.
\end{proof}

{\bf Proof of part $(ii)$ of Theorem \ref{fin}.} Let the conditions
in part $(ii)$ of Theorem \ref{fin} be fulfilled. Then the equality
\begin{equation}\label{Slambda}
\lim\limits_{|z-m|\to 0} |log|z-m||^{-1}n(1,T(z)) ={U}(1)
\end{equation}
follows from Lemmas \ref{H-SH},\,\ref{comp.pert} and Theorem
\ref{main}. Taking into account the equality \eqref{Slambda} and
using Lemma \ref{b-s} we complete the proof of part $(ii)$ of
Theorem \ref{fin}. \qed

{\bf Acknowledgement} The authors would like to thank
Prof.~R.~A.~Minlos  for helpful discussions about
 the results of this paper. This work was
supported by the DFG 436 USB 113/4 Project and the Fundamental
Science Foundation of Uzbekistan. S.~N.~Lakaev and Z.~I.~Muminov
gratefully acknowledge the hospitality of the Institute of Applied
Mathematics and of the IZKS of the University of Bonn.

\end{document}